\documentclass[doublecol]{epl2}
\usepackage{amssymb}

\title{Influence of spin polarization on resistivity of a two-dimensional electron
gas in Si MOSFET at metallic densities}
\shorttitle{In-plane magnetoresistance in a metallic MOSFET 2DEG}

\author{I. Shlimak\inst{1}\footnote{E-mail: shlimai@biu.ac.il}, A. Butenko\inst{1}, D. I. Golosov\inst{1}, K.-J.
Friedland\inst{2}, S. V. Kravchenko\inst{3}}
\shortauthor{I. Shlimak \etal}

\institute{
   \inst{1} Jack and Pearl Resnick Institute of Advanced Technology,
Department of Physics, Bar-Ilan University, Ramat-Gan 52900, Israel\\
   \inst{2} Paul-Drude Institut f\"{u}r Festk\"{o}rperelektronik,
Hausvogteiplatz 5-7, 10117, Berlin, Germany\\
   \inst{3} Physics Department, Northeastern University, Boston, MA 02115,
U.S.A.}


\abstract{Positive magnetoresistance (PMR) of a silicon MOSFET in
parallel magnetic fields $B$ has been measured at high electron
densities $n \gg n_{\rm c}$ where $n_{\rm c}$ is the critical density
of the
metal-insulator transition (MIT). It turns out that the normalized
PMR curves, $R(B)/R(0)$, merge together when the field is scaled
according to $B/B_{\rm c}(n)$ where $B_{\rm c}$ is the field in which
electrons become fully spin polarized. The values of $B_{\rm c}$ have
been calculated from the simple equality between the Zeeman splitting
energy and the Fermi energy taking into account the experimentally
measured dependence of the spin susceptibility on the
electron density. This extends the range of validity of the scaling
all the way to a deeply
metallic regime far away from MIT.
 The subsequent analysis of PMR for low  $n\gtrsim n_{\rm c}$
demonstrated that
 the merging 
of the initial parts of curves 
can bee achieved only with 
taking into account the temperature dependence of $B_{\rm c}$. It is 
also shown that the shape of the PMR curves at strong magnetic fields is affected 
by a  crossover from a purely two-dimensional (2D) electron
transport to a regime where out-of-plane carrier motion becomes important
(quasi-three-dimensional regime).}

\begin{document}

\maketitle

\section{Introduction}

Following the experimental observation of the the
metal-insulator transition (MIT) in two-dimensional (2D)
systems (for a review, see, for example, Ref.~\cite{spivak10}), the
resistivity of various 2D systems, including silicon-based
metal-oxide-semiconductor field-effect transistors (MOSFETs) and GaAs-based
heterostructures, was measured in a parallel magnetic field in the vicinity
of MIT~\cite{simonian97,Pudalov, Okamoto, Mertes, Vitkalov, ShashKravch,
ShashDev, Gao}. These studies were largely motivated by the still open
question concerning the influence of the electron spin polarisation on the
MIT in 2D (indeed, a parallel magnetic field lifts the spin degeneracy
without otherwise influencing the electron motion).

It has been observed that on the metallic side of the MIT, where $n>n_{%
\mathrm{c}}$ (for Si MOSFETs, $n_{\mathrm{c}}\approx 0.8\div 1.0\cdot
10^{11}~\mathrm{cm}^{-2}$, depending on the degree of disorder), the sample
resistance $R$ increases with the parallel magnetic field, saturating above
some characteristic value of the field, $B_{\mathrm{c}}$. This value is
interpreted as the field corresponding to
complete spin polarisation of
the electrons. Possible mechanisms of the influence of the spin polarisation
on the resistance were discussed theoretically in Refs. \cite{
spivak10,
Dolgopolov,
Zala}.

It was also observed \cite{Vitkalov, ShashKravch} that once the resistivity
and magnetic field are rescaled as $R(B,n)/R(0,n)$ and $B/B_{\mathrm{c}}(n)$%
, the PMR curves measured at different densities merge into one. This
enables one to find the values of $B_{\mathrm{c}}$ for higher electron
densities where the full spin polarisation cannot be experimentally
achieved: $B_{\mathrm{c}}$ can be considered as a variable parameter, to be
determined from the best merging. Below, we will use this prescription.

In turn, the dependence of $B_{\mathrm{c}}$ on electron density has
attracted significant attention. In Ref.~\cite{ShashKravch}, it was shown
that $B_{\mathrm{c}}$ is an almost linear function of $n$ vanishing at a
sample-independent value $n=n_{\chi }\approx 0.8\cdot 10^{11}~\mathrm{cm}%
^{-2}$. This tendency of $B_{\mathrm{c}}$ to vanish at a finite electron
density raises the question about a possible magnetic instability in a
strongly correlated 2DEG, resulting in a spontaneous spin polarisation at $%
n<n_{\chi }$. Then the 2DEG magnetic susceptibility $\chi $, which is
proportional to the product $g^{\ast }m^{\ast }$ of the effective Land\'{e} $%
g$-factor and the renormalised effective mass $m^{\ast }$, must be singular
(divergent) at $n=n_{\chi }$. Interestingly, critical 
density
of
suggested ferromagnetic instability $n_{\chi }$ as found in Ref.~\cite
{ShashKravch} for a (100) Si-MOSFET \footnote{%
A strong enhancement of $m^{*}$ with decreasing electron density $n$ was
also observed in a (111) Si-MOSFET\cite{Shashkin2007}. Analysis of these
data shows \cite{Gold2007} that the behaviour of $m^{*}(n)$ is similar to
that in a (100) Si-MOSFET.} is close to $n_{\mathrm{c}}=0.8\cdot 10^{11}~%
\mathrm{cm}^{-2}$, the latter corresponding to the MIT. On the other hand,
according to Ref.~\cite{PudalGersh}, the observed gradual increase of the
product $g^{\ast }m^{\ast }$ 
upon
 approaching $n_{\mathrm{c}}$ from the
metallic side does not lead to an actual divergence of $\chi $ at $n_{%
\mathrm{c}}$. This disagreement is discussed in Refs. \cite{Kravchenko,
Fleury}. Thus, the dependence $B_{\mathrm{c}}(n)$ near $n_{c}$ remains an
open question in the problem of the influence of electron spin polarisation
on MIT in 2D.

In the vicinity of $n_{\mathrm{c}}$, the correct experimental values of $B_{%
\mathrm{c}}$ can be obtained only via very low temperature measurements
because $B_{\mathrm{c}}$ shows a strong temperature dependence \cite{Gao}.
Meanwhile, far from the MIT, at $n\gg n_{\mathrm{c}}$, 
the degree of the spin polarisation is
almost temperature-independent. In this region, the value of $B_{\mathrm{c}}$
can be determined from the ratio between Zeeman splitting 
$\Delta \varepsilon _{%
\mathrm{Z}}=g^{\ast }\mu _{\mathrm{B}}B$ (where $\mu _{\mathrm{B}}$ is
the Bohr magneton) and the Fermi energy $\varepsilon _{\mathrm{F}}=
2 n(\pi
\hbar ^{2}/m^{\ast }\nu _{\mathrm{s}}\nu _{\mathrm{v}})$, where $\nu _{%
\mathrm{s}}$ and $\nu _{\mathrm{v}}$ are the spin and valley degeneracy
factors. As a result, the experimental values of $B_{\mathrm{c}}$ obtained
at low temperatures correctly represent the $T=0$ case. At these high
densities, the 2DEG physics should be described by a conventional Fermi
liquid picture; it is therefore instructive to compare it to the results
obtained at lower densities where the complications related to the proximity
to the MIT become pronounced. 

In this Letter, we report measurements of PMR at high electron densities $%
n\gg n_{c}$; the subsequent analysis allows for determination of $B_{c}(n)$
and for a meaningful comparison with the low-density case $n\gtrsim n_{c}$.

\section{Experimental}

We measure the resistivity of a Si MOSFET sample in a magnetic field
parallel to the surface and directed along the current. The slot-gate sample
was similar to those described in our previous publications \cite{ShlimGin,
ShlimGol}. The electron density was controlled by the gate voltage $V_{%
\mathrm{G}}$: $n = 1.24\cdot 10^{11}~\mathrm{cm}^{-2}(V_{\mathrm{G}} - 0.34%
\mathrm{V})$. The PMR curves were measured at $T = 0.3$~K by the standard
lock-in technique.

The high sensitivity of our setup allows for measuring a weak PMR signal at
high electron densities up to $n=2.1\cdot 10^{12}\mathrm{cm}^{-2}$. The
actual values of gate voltages applied (and hence the densities achieved)
are restricted by the gate breakdown at high $V_{G}$.

The data used in Figs.~\ref{fig:4} and \ref{fig:5} were obtained in an
earlier measurement at $T=$ 50 mK, partly reported in Ref. \cite{ShashKravch}.


\begin{figure}[tbp]
\onefigure[width=8cm]{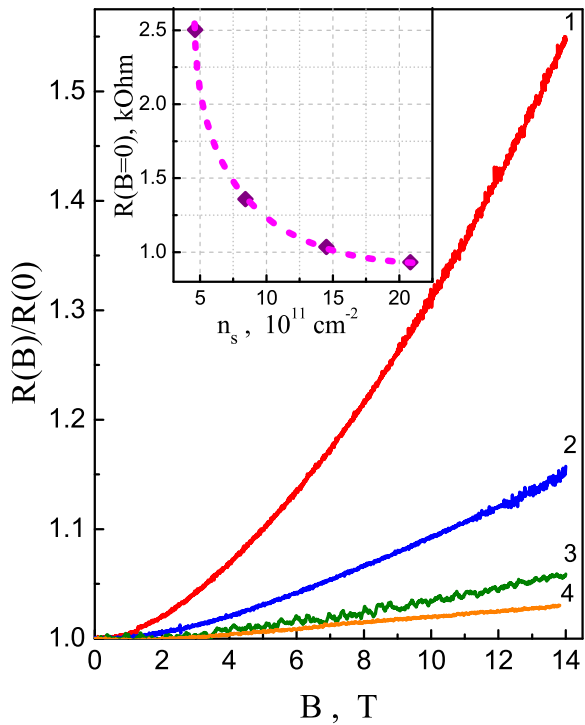}
\caption{(colour online) Normalised magnetoresistance $R(B)/R(0)$ of a Si MOSFET in
a parallel magnetic field. Curves 1 to 4 correspond to gate voltages $V_G=4 
\mathrm{V}$, $V_G=7 \mathrm{V}$, $V_G=12 \mathrm{V}$, and $V_G=17 \mathrm{V}
$, respectively. Resistance $R(0)$ in the absence of the field is shown
in the inset.}
\label{fig:1}
\end{figure}

\section{Results and discussion}

Fig.~\ref{fig:1} shows the PMR curves $R(B)/R(0)$ measured at different
values $V_{\mathrm{G}}$ and, hence, at different $n$. One can see that the
PMR decreases rapidly with increasing density (the corresponding values of $%
R(0)$ as a function of $n$ are shown in the inset to Fig.~1). The data allow
us to calculate $B_{\mathrm{c}}$ from the condition that the Zeeman 
splitting and the
Fermi energy are equal: 
\begin{equation}
B_{\mathrm{c}}=\frac{\pi \hbar ^{2}n}{(g^{\ast }m^{\ast })\mu _{%
\mathrm{B}}}.  \label{eq:zeefermi}
\end{equation}

When calculating the Fermi energy, it is taken into account that $\nu _{%
\mathrm{v}}=2$ for Si(100) MOSFET, and that $\nu _{\mathrm{s}}=1$ in a
strong magnetic field. The other parameter that enters Eq. (\ref{eq:zeefermi}%
) is the renormalised value $\chi \propto (g^{\ast }m^{\ast })$ which
depends on the density. This latter dependence was reported in Ref.~\cite
{PudalGersh} in the dimensionless scale $(g^{\ast }m^{\ast })/2m_{\mathrm{b}}
$ vs. $r_{\mathrm{s}}$. Here, 2 is the bare $g$-factor for Si, $m_{\mathrm{b}%
}=0.19m_{0}$ the ``bare'' effective mass in Si (100), and $r_{\mathrm{s}}$
is the ratio between the Coulomb interaction energy and Fermi energy, $r_{%
\mathrm{s}}=m_{\mathrm{b}}e^{2}/4\pi \epsilon \epsilon _{0}\hbar (\pi
n)^{1/2}$ ( with $\epsilon =7.7$ the dielectric constant of the interface
between Si and SiO$_{2},$ $\epsilon _{0}$ is vacuum permittivity). For Si
MOSFET, one finds $r_{\mathrm{s}}=2.65\cdot (10^{12}~\mathrm{cm}%
^{-2}/n)^{1/2}$. The calculated values of $B_{\mathrm{c}}$ are shown in
Table \ref{tab:bc}. \renewcommand{\arraystretch}{1.5} 
\begin{table}[tbp]
\caption{Calculated values of $B_{c}$. The dependence of $g^{\ast }m^{\ast
}/2m_{b}$ on $r_{s}$ is taken from Ref. \protect\cite{PudalGersh}. }
\label{tab:bc}
\begin{center}
\begin{tabular}{cccccc}
\hline
No. & $V_{\mathrm{G}}$, V & $n$, $10^{12} \mathrm{cm}^{-2}$ & $r_{\mathrm{s}}$ & $%
g^*m^*/2m_{\mathrm{b}} $ & $B_{\mathrm{c}}$, T \\ \hline\hline
1 & 4 & 0.46 & 3.91 & 1.89 & 24.5 \\ 
2 & 7 & 0.84 & 2.89 & 1.64 & 41.4 \\ 
3 & 12 & 1.45 & 2.20 & 1.47 & 98.8 \\ 
4 & 17 & 2.08 & 1.84 & 1.38 & 151 \\ \hline
\end{tabular}
\end{center}
\end{table}

Because of the high electron densities, we are far from the saturation
regime, and the values of $B_{\mathrm{c}}$ can be tested only by merging all
the PMR curves together by scaling them with $B/B_{\mathrm{c}}$. 
\begin{figure}[tbp]
\onefigure[width=8cm]{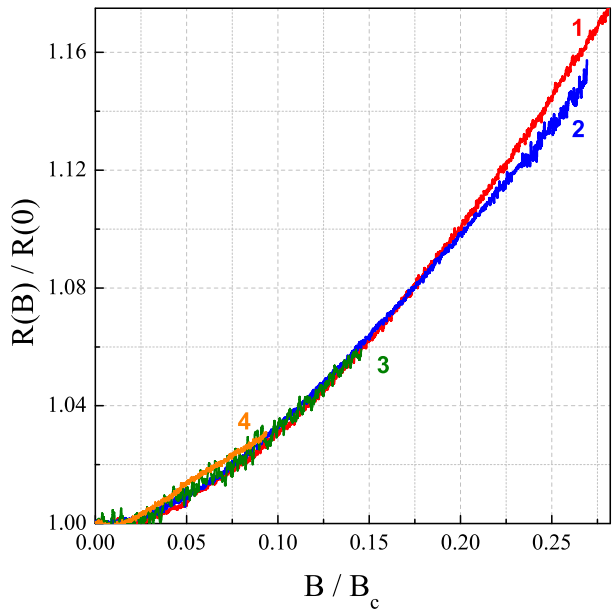}
\caption{(colour online) Normalised magnetoresistance rescaled as a function of $%
B/B_{c}$. Curves 1 to 4 correspond to gate voltages $4\mathrm{V}$, $7\mathrm{%
V}$, $12\mathrm{V}$, and $17\mathrm{V}$. The position of each curve number on
the plot marks the end of the corresponding curve.}
\label{fig:2}
\end{figure}
Fig.~\ref{fig:2} shows the result of this scaling. One can see that a good
merging of all the curves supports our calculated values of $B_{\mathrm{c}}$. 
Finally, in Fig.~\ref{fig:3} we plot the dependence of $B_{\mathrm{c}}(n)$. 
We note that at lower densities, the values of $B_{\mathrm{c}}(n)$
obtained from Eq.~(\ref{eq:zeefermi}) are close to those determined directly
from the saturation of the PMR at high fields \cite{Pudalov,ShashKravch,Broto}.

\begin{figure}[tbp]
\onefigure[width=8cm]{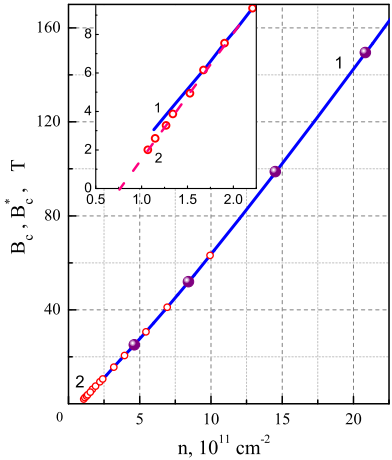}
\caption{(colour online) $B_\mathrm{c}$ as a function of $n$, Eq. 
(\ref{eq:zeefermi}), for high electron densities. Values given in 
Table \ref{tab:bc} (and used
in Fig.~\ref{fig:2}) are shown by bullets. The inset shows a blow-up of the
low-density region.}
\label{fig:3}
\end{figure}

\begin{figure}[tbp]
\onefigure[width=8cm]{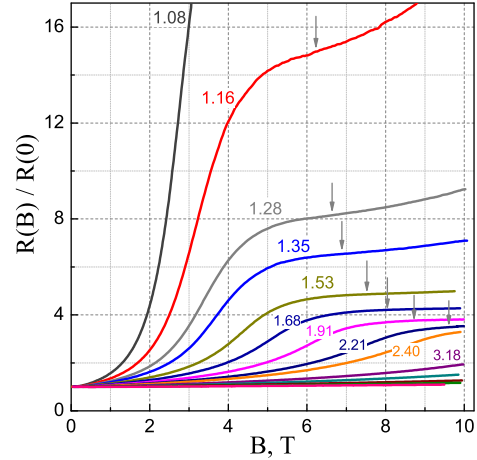}
\caption{(colour online) Magnetic field dependence of normalised magnetoresistance
in the region of lower carrier densities (marked in the units of $10^{11} 
\mathrm{cm}^{-2}$). Arrows denote the respective values of $B_{2D}$ 
calculated 
for $\alpha = 0.45$ as
discussed in the text.}
\label{fig:4}
\end{figure}

\begin{figure}[tbp]
\onefigure[width=8cm]{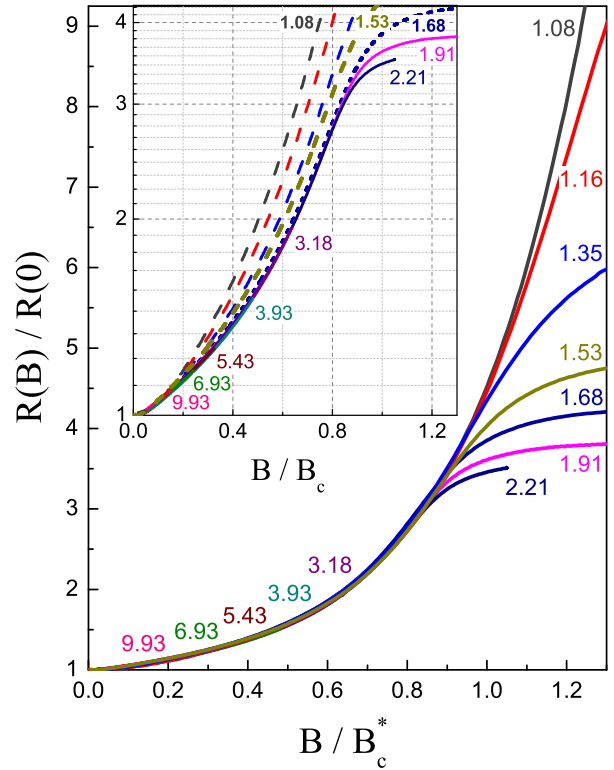}
\caption{(colour online) Normalised magnetoresistance $R(B)/R(0)$
scaled as a function of $B/B_{\rm c}$ (inset) and
as a function of $B/B_{\rm c}^*$ (main panel). Here, $B_{\rm c}$ is given 
by eq.  (\ref{eq:zeefermi}). 
and $B^*_{\rm c}$ is discussed in the text. Values of electron density
$n$ are in  units of $10^{11}\mathrm{cm}^{-2}$; 
for $n\geq 2.21\mathrm{cm}^{-2}$, the position of each value on the plot
marks the end of the corresponding curve.}
\label{fig:5}
\end{figure}

We will now discuss the validity of the Fermi liquid-like approach which
underlies Eq.~(\ref{eq:zeefermi}) at these lower densities. Fig.~\ref{fig:4} 
shows the
PMR dependencies measured for the low electron densities $n\gtrsim n_{c}$ ($1
\div 3 \cdot 10^{11}{\rm cm}^{-2}$) at $T=$ 50mK and partly published in Ref.
\cite
{ShashKravch}.  In Fig. \ref{fig:5}, we plot these PMR curves $R(B)/R(0)$ as 
functions of $B/B_{\mathrm{c}}$, where $B_{\mathrm{c}}(n)$ is obtained from
Eq.~(\ref{eq:zeefermi}) with appropriately renormalised\cite{PudalGersh}
values of $(g^{\ast }m^{\ast })$. The result is shown in Fig.\ref{fig:5} 
in the inset. We see that at the densities below $n=1.68 \cdot 10^{11} {\rm cm}^{-2}$, the PMR curves at low fields deviate from the common scaled curve
describing the higher-density regime.
We thus conclude that at these lower densities, our calculation of 
$B\mathrm{_{c}}(n)$ [see Eq. (\ref{eq:zeefermi})]
becomes inaccurate. Indeed, the data for $g^\ast m^{\ast }$ \cite
{PudalGersh} used in our Fig.~\ref{fig:3} were taken at 300 mK, while the MR
curves were measured at 50 mK. It is known\cite{Gao} that in the region near
MIT the value of $B_{\rm c}$ (or equivalently $g^{\ast }m^{\ast }$) 
shows strong
temperature dependence. Hence the data of Ref.\cite{PudalGersh} might not be
appropriate for the analysis of our $R(B)$ curves at lower $n$, taken at
about 50 mK. However, the higher-density regime, where Eq. (\ref{eq:zeefermi})
works perfectly, provides a reliable reference point for evaluating $B_c^*(n)$ 
from
the requirement that the initial parts of all the $R(B/B_c^*)/R(0)$ collapse.
The values of $B_c^*$, obtained in this way, are shown in Fig. \ref{fig:3}
in the inset
(dashed line). One sees that the values of $B_c^*$ indeed 
deviate from
that of $B_c$ (solid line) at smaller $n$. The PMR curve merging is
illustrated in Fig.~\ref{fig:5} (main panel). One may
conclude that for the lower concentrations, $n\stackrel{<}{\sim }
1.68 \cdot 10^{11} {\rm cm}^{-2}$, the
value of $k_{B}T$ with $T=300$ mK becomes smaller that the energy scale of
the critical fluctuations, and therefore the values of $g^{\ast }m^{\ast }$
obtained in Ref.~\cite{PudalGersh} no longer correspond to the true
low-temperature limiting case (cf. Ref.~\cite{Kravchenko}).
Nevertheless, we note that in Ref.\cite{PudalGersh}, the data for 
$g^\ast m^{\ast }$ were obtained from the 
analysis of Shubnikov--de Haas oscillations, whereas in our Fig. ~\ref{fig:3}
the values of $B_c^\ast$ (suggesting slightly different $g^\ast m^{\ast }$) 
were determined using an entirely different method, viz., the merging of all 
the MR curves. The absence of a precise agreement between the two is 
therefore not surprising.

It is seen from Fig.~\ref{fig:4} that  down 
to $n=1.68\cdot 10^{11}\mathrm{cm}^{-2}$,
MR curves are saturated above $B=B_{\mathrm{c}}$, which shows that the value
of $B_{\mathrm{c}}$ indeed corresponds to the complete spin polarisation.
The behaviour at lower values of $n$ is somewhat different. We note that at 
$n<1.53\cdot 10^{11}{\rm cm}^{-2}$, the MR curves at $B=B_{\mathrm{c}}$ exhibit
only a bend followed by a further increase of resistance with increasing
magnetic field, and this behaviour becomes more pronounced with decreasing $n$.
This feature is not related to the peculiar physics underlying
the saturation of resistivity at $B=B_{\mathrm{c}}$ as outlined above. Rather,
we find that this further increase of resistance reflects a crossover 
from 2D to a quasi-3D mechanism of 
conductivity,  due to the carrier motion in the out-of-plane direction which
begins to 
play a role in a strong magnetic field, $B\stackrel{>}{\sim }B_{2D}$.
At $B= B_{2D}$,  magnetic
length $\lambda =(\hslash /eB)^{1/2}$ becomes comparable to the thickness of 2D
layer $Z$, viz. $\lambda = \alpha Z$.
Here, $\alpha$ is a pre-factor of the order of unity, to be determined
phenomenologically; we find that $\alpha = 0.4 \div 0.5$ yields a
good fit (see the arrows in Fig. \ref{fig:4}).

\begin{figure}[tbp]
\onefigure[width=8cm]{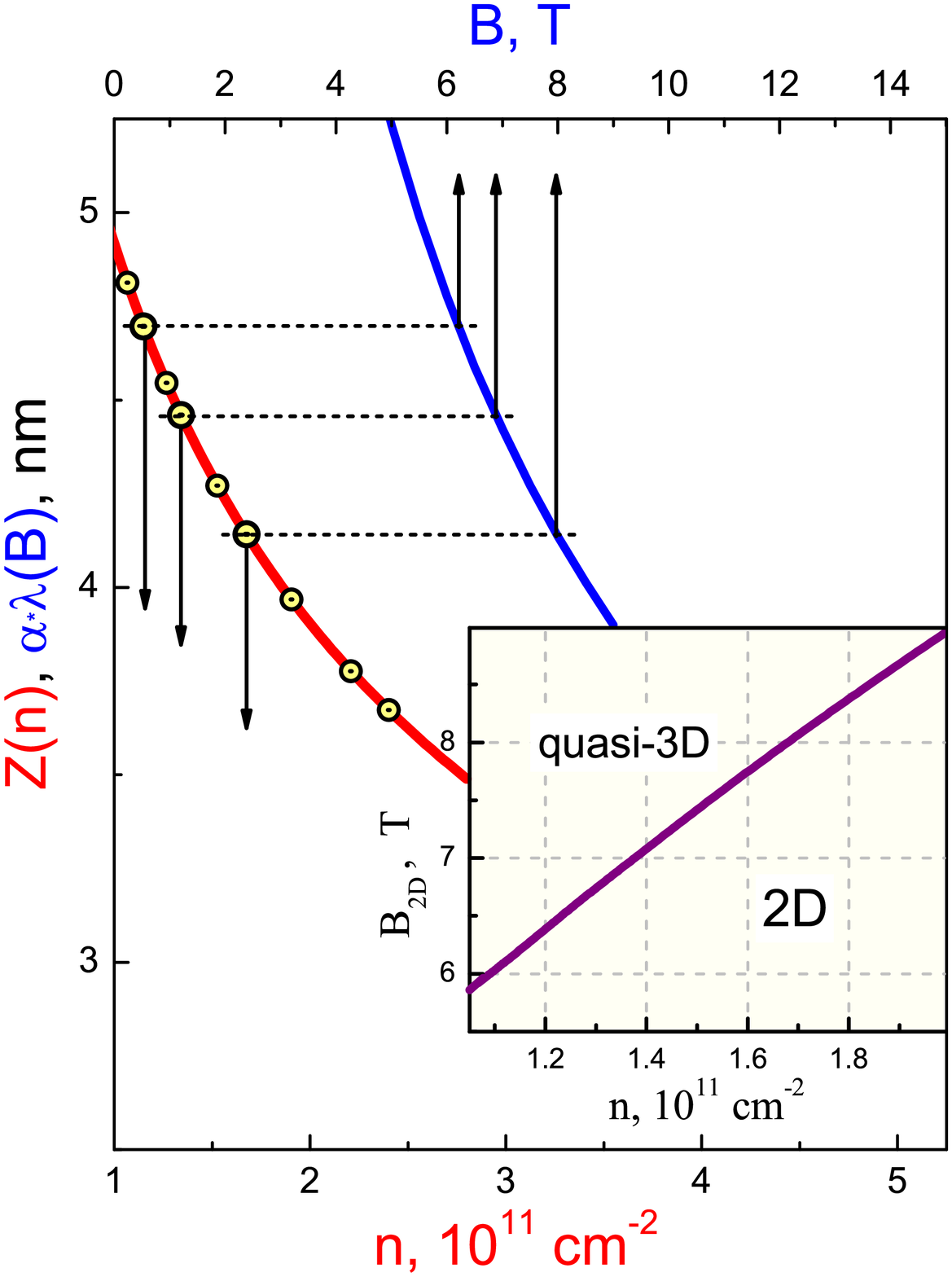}
\caption{(colour online) The dependence of conducting layer thickness $Z$ on
the electron density $n$ (lower scale), and the scaled magnetic 
length $\alpha \lambda$ (for $\alpha = 0.45$) as a function of the 
magnetic field (upper scale). The resultant
dependence of $B_{2D}$ on $n$ is shown in the inset.}
\label{fig:6}
\end{figure}

The values of $Z(n)$ (as estimated in Ref. \cite{Ando})  and 
of $\lambda(B)$  
are plotted in Fig. \ref{fig:6} (lower and upper curves, respectively).
The dependence of $B_{2D}$ on $n$ is  shown in the inset. 
One can see that
within the experimental interval of magnetic fields (up to 14 T),
conductivity of samples with higher $n$ always has a purely 2D 
character. However
for smaller $n$, when the corresponding $B_{2D}$ is below 14T, the measured
conductivity exhibits a transition from 2D to quasi-3D character with increase of $%
B$. This explains the absence of saturation and further increase of sample
resistance $R$ after the bend in Fig.~4. Similar MR behaviour had been
observed in bulk (3D) heavily doped Si with metallic conductivity\cite
{Shafarman}. 

\section{Conclusion}

When viewed in the context of the earlier work on the lower-density Si
MOSFETs, our results suggest the following generic conclusions:

(i) We have found that the PMR is present in the high-density 2DEG far away
from the MIT. In this region, the system is essentially a Fermi-liquid and
its transport properties (including the PMR) are not directly related to the
occurrence of the MIT at lower densities. This provides a further support to
the theoretical studies \cite{Zala,Dolgopolov} which obtain the PMR based on
the Fermi liquid description. We speculate that a related magnetic field
dependence would likely arise in the diffusion coefficient
of a 2DEG without impurities. 

(ii) We emphasise nevertheless that the PMR observed in the high-density
regime 
must have
 the same underlying physics as the much stronger effect observed
closer to the MIT. This is unequivocally illustrated by the fact that the
same dimensionless scaling (in terms of $B/B_{c}(n)$) works throughout both
regimes, which connect seamlessly. The scaling reflects a Fermi liquid-type
renormalisation of the effective 2DEG parameters (in this case, $g^{\ast }$
and $m^{\ast }$). This suggests that the Fermi-liquid phenomenology itself
is valid all the way down to the vicinity of the MIT, as long as the system
remains in the ballistic regime. At yet lower densities, in the diffusive
regime, the non-trivial interplay between the interactions and disorder,
much discussed in the literature (see, \textit{e.g.}, Ref.\cite{punnoose05}%
), strongly affects the Fermi liquid parameters, resulting in a strong
increase of the spin susceptibility with decreasing temperature, but this is
beyond the scope of the present paper.

(iii) We find that with an increase of the parallel magnetic field $B$
beyond a certain value $B_{2D}$ at a fixed $n$, the character of
conductivity undergoes a crossover from  two-dimensional  to a
quasi-three-dimensional 
one
when the magnetic length $\lambda $ becomes small 
in comparison to
the layer width $Z$. This is directly observed in the regime of smaller
densities $n$, where the value of $B_{2D}$ lies within the 
experimentally
studied range of $B$ up to 14 T. The quasi-3D behaviour at high fields is reflected
in a continued increase of sample resistance with $B$ 
instead of saturation.






\acknowledgments
We take pleasure in thanking A. Belostotsky and T. Hasenfratz for assistance,
and R. Berkovits for helpful discussions.
This work was supported in part by the BSF Grant 2006375 and by the Israeli
Ministry of Immigrant Absorption. 
SVK was supported by DOE Grant DE-FG02-84ER45153. IS
thanks the Erick and Sheila Samson Chair of Semiconducting Technology for
financial support.

\end{document}